\documentclass{PoS}
\usepackage{epsfig}
\usepackage{cite}
\newcommand{\postscript}[2]{\setlength{\epsfxsize}{#2\hsize}
   \centerline{\epsfbox{#1}}}
\newcommand{\comment}[1]{}

\usepackage[usenames,dvipsnames]{xcolor}
\definecolor{orange}{cmyk}{0,0.5,1,0}
\definecolor{rossoCP3}{cmyk}{0,.88,.77,.40}
\definecolor{graa}{rgb}{0.8,0.8,0.8}
\definecolor{blaa}{rgb}{0.2,0.2,0.6}

\title{Probing QCD approach to thermal equilibrium with ultrahigh energy cosmic rays}

\ShortTitle{Probing QCD approach to thermal equilibrium with UHECR}

\author{\speaker{Jorge F. Soriano} \\
Department of Physics \& Astronomy,  Lehman College, CUNY, NY 10468, USA\\
Department of Physics,
 Graduate Center, City University
  of New York,  NY 10016, USA\\
E-mail: \email{jfdezsoriano@gmail.com}}

\author{Luis A. Anchordoqui\\
Department of Physics \& Astronomy,  Lehman College, CUNY, NY 10468, USA\\
Department of Physics,
 Graduate Center, City University
  of New York,  NY 10016, USA\\
Department of Astrophysics,
 American Museum of Natural History, NY
 10024, USA\\
        E-mail: \email{luis.anchordoqui@gmail.com}}

\author{Thomas C. Paul\\
Department of Physics \& Astronomy,  Lehman College, CUNY, NY 10468, USA\\
E-mail:\email{tompaulster@gmail.com}}

\author{Thomas J. Weiler\\
 Department of Physics \& Astronomy, Vanderbilt University, Nashville, TN 37235, USA\\
        E-mail: \email{tom.weiler@vanderbilt.edu}}

      \abstract{The Pierre Auger Collaboration has reported an excess
        in the number of muons of a few tens of percent over
        expectations computed using extrapolation of hadronic
        interaction models tuned to accommodate LHC data. Very
        recently, we proposed an explanation for the muon excess
        assuming the formation of a deconfined quark matter (fireball)
        state in central collisions of ultrarelativistic cosmic rays
        with air nuclei. At the first stage of its evolution the
        fireball contains gluons as well as $u$ and $d$ quarks. The
        very high baryochemical potential inhibits gluons from
        fragmenting into $u \bar u$ and $d \bar d$, and so they
        fragment predominantly into $s \bar s$ pairs. In the
        hadronization which follows this leads to the strong
        suppression of pions and hence photons, but allows heavy
        hadrons to be emitted carrying away strangeness. In this
        manner, the  extreme imbalance of hadron to photon content
        provides a way to enhance the muon content of the air
        shower. In this communication we study theoretical systematics
        from hadronic interaction models used to describe the cascades
        of secondary particles produced in the fireball explosion. We
        study the predictions of one of the leading LHC-tuned models
        QGSJET II-04 considered in the Auger analysis.}

\FullConference{35th International Cosmic Ray Conference - ICRC217 -\\
		10-20 July, 2017\\
		Bexco, Busan, Korea}

\begin{document}

Besides addressing central questions in ultrahigh energy cosmic ray
(UHECR) astrophysics -- determining the baryonic component and
identifying or constraining sources -- the AugerPrime upgrade will
provide unique access to particle physics at an order-of-magnitude
higher center-of-mass energy than the Large Hadron Collider
(LHC)~\cite{Aab:2016vlz}.

As recently demonstrated by the Pierre Auger
Collaboration~\cite{Aab:2016hkv}, it is possible to test particle
physics well above 100~TeV in the UHECR-air nucleon center-of-mass
energy, using hybrid UHECR air showers, even with a mixed primary
composition.  Moreover, the column energy-density in UHECR-air
collisions is an order of magnitude greater than in Pb-Pb collisions
at the LHC, suggesting the potential for new hadronic physics from
gluon saturation and possibility of exploring quark-gluon plasma as
well as quark matter formation by heavy nuclear primaries at far higher energies than
available in accelerators~\cite{Farrar:2013sfa,Anchordoqui:2016oxy}.

A significant discrepancy in the shower muon content is found
(greater than $2\sigma$, statistical and systematics combined in
quadrature) between predictions of LHC-tuned hadronic event generators
and observations~\cite{Aab:2016hkv}.  With the added
muon-electromagnetic separation and the significantly higher
data-taking rate for the highest energy hybrid events provided by
AugerPrime, the reason for this discrepancy  may be determined.

As a UHECR-induced air shower develops in the atmosphere, it increases
in particle number, before eventually the particle energies drop below
some threshold, at which ionization losses begin to cull the particle
population.  The position at which an air shower deposits the maximum
energy per unit mass traversed is known as $X_{\rm max}$ and its
dispersion is known as $\sigma (X_{\max})$. Both of these observables are
sensitive to the primary UHECR composition, though interpreting this
dependence requires resorting to predictions of simulations of
hadronic interactions at energies and in kinematical regions
inaccessible to terrestrial experiments~\cite{Anchordoqui:2004xb}.

The Auger surface array can also be employed to extract
composition-dependent information, including the shower muon richness
observed a the ground as well as the muon production depth, $X_{\rm
  nax}^\mu$, which is the depth along the shower axis where muon
production reaches a maximum.  One advantage of the surface array is
its nearly 100\% duty cycle, compared to the roughly 15\% duty cycle
of the fluorescence detectors.  A measure of $X_{\rm max}^\mu$ can be
attained via timing information of the muons when they arrive at the
ground.  Curiously, the primary masses predicted by $X_{\rm max}$ and
$X_{\rm max}^\mu$ are not in agreement, with the $X_{\rm max}$
suggesting an average composition at the highest energies in the mass
range of nitrogen~\cite{Aab:2014kda}, while $X_{\rm max}^\mu$ is more
consistent with a composition heavier than iron, using LHC-tuned EPOS
at the hadronic interaction model~\cite{Aab:2014dua}. Using QGSJET
leads to a better agreement between $X_{\rm max}$ and $X_{\rm
  max}^\mu$, with $X_{\rm max}^\mu$ favoring iron composition at the
highest energies. Regardless of models chosen, however, it appears
difficult at present to reconcile the $X_{\rm max}$ and $X_{\rm
  max}^\mu$ measurements with each other using existing hadronic
interaction models.

In addition, it is interesting to note that while for $10^{9.5}
\lesssim E/{\rm GeV} \lesssim 10^{10.6}$ the mean and dispersion of
$X_{\rm max}$ inferred from fluorescence Auger data point to a light
composition (protons and helium) towards the low end of this energy
bin and to a large light-nuclei content (around helium) towards the
high end (see Fig.~3 in~\cite{Aab:2016zth}), when the signal in the
water Cherenkov stations (with sensitivity to both the electromagnetic
and muonic components) is correlated with the fluorescence data, a
light composition made up of only proton and helium becomes
inconsistent with observations~\cite{Aab:2016htd}. The hybrid data indicate that
intermediate nuclei, with baryon number $A = 14$, must contribute to
the energy spectrum in this energy bin. Moreover, a potential iron
contribution cannot be discarded.

New physics processes in the first interaction would tend to increase
$\sigma (X_{\rm max})$, making the nuclear composition appear lighter
than what actually is. This is because we would not expect new physics
processes when nuclei just slide along each other. The admixture of
peripheral and ``new physics'' collisions would then produce large
fluctuations in the number of muons at ground level and increase
$\sigma (X_{\rm max})$. On the other hand, $X_{\rm max}^\mu$ occurs
after the shower is more fully thermalized and hence $\sigma (X_{\rm
  max}^\mu)$ is less sensitive to new physics in the first
interaction.  Combining information from these two variables thus
provides a means of disentangling effects of nuclear composition from
new physics. If measurements of $\sigma (X_{\rm max})$ and $X_{\rm
  max}^\mu$ were equally accessible experimentally, $X_{\rm max}^\mu$
would provide a better measurement of the nuclear composition, as it is less
sensitive to new physics in the first interaction.

It is also important to note that any new physics should affect
$\sigma (X_{\rm nax})$, causing the distribution to trend upwards
above the new physics threshold, an effect which would be too large to
be accounted for by increasing primary mass. On the other hand, new
physics would not manifest in this way in the case of $X_{\rm max}$.
A constant nuclear composition would yield a straight elongation rate,
while an increase in primary mass would cause $X_{\rm max}$ to inflect
downward. These unique predictions for $X_{\rm max}$ and $\sigma
(X_{\rm max})$, are in agreement with existing data; see again Fig.~3
in~\cite{Aab:2016zth}.  With enough statistics, it should be possible
to discern these effects.

Very recently we presented a model that can accommodate all these
anomalies~\cite{Anchordoqui:2016oxy}. The model builds up on an old
idea, which allows formation of a deconfined quark matter (fireball)
state in central collisions of ultrarelativistic cosmic rays with air
nuclei~\cite{Halzen:1981zx}. At the first stage of its evolution the fireball contains
gluons as well as $u$ and $d$ quarks. The very high baryochemical
potential inhibits gluons from fragmenting into $u \bar u$ and $d \bar d$,
and so they fragment predominantly into $s \bar s$ pairs. In the
hadronization which follows this leads to the strong suppression of
pions and hence photons, but allows heavy hadrons to be emitted
carrying away strangeness. In this manner, the extreme imbalance of
hadron to photon content provides a way to enhance the muon content of
the air shower. In addition, the admixture of peripheral and
fireball collisions would then produce large fluctuations on $X_{\rm
  max}$, but the muon shower maximum $X_{\rm max}^\mu$ would have
small fluctuations, in agreement with existing data.

In this paper we use simulations to study the effects of fireball models in air
shower observables. Cosmic Ray Monte Carlo (CRMC v1.6.0)~\cite{Baus:2016} is an
interface to different hadronic event generators such as EPOS
LHC~\cite{Pierog:2015} or QGSJET II-04~\cite{Ostapchenko:2011}. It outputs
the secondaries of particle interactions for different hadronic
models, which allows an standard use of the interaction products for any further
analysis. CORSIKA (v7.5600)\cite{Heck:1998} performs the simulation of atmospheric
showers produced after cosmic ray interactions in the atmosphere. It makes use
of different hadronic event generators to obtain the products of the first and
subsequent interactions at the highest energies. Nevertheless, the first
interaction can be treated externally (STACKIN option), allowing the user to
give as input not just the cosmic ray primary, but the secondaries and their energies
and momenta. It is possible then to modify the particle content and properties
after the first interaction to analyze afterwards the evolution of the shower.

Our goal is to test whether fireball-like models would be able to mitigate the
current tension between LHC extrapolated predictions and data at the Pierre
Auger Observatory, without developing a full theoretical framework for fireball
models or carrying out fully detailed simulations of the physical processes
involved in fireball interaction secondary particle production. In this work we
take CRMC results for first interactions and modify the particle content in a
consistent manner according to the above description of the fireball model,
where the pion content is highly suppressed. We study the effect of this
modification on the muon content at the Pierre Auger observation level.

The computational power needed to run full simulations of atmospheric showers is
an important limiting factor in the accuracy that one can reach in simulated air
shower analyses. The thinning option in CORSIKA allows a reduction of the
computing time by reducing some information about the development of individual
particles and compensating the by applying weights to simulated particles. A
comparison of results for thinned and unthinned showers presented in
\cite{Pierog:2015icrc} allows us to consider that, when the expected number of
particles at observation level is high enough, the fluctuations (of the number
of particles) created by the thinning algorithm are small enough for the result
to be close to the corresponding unthinned result. Nevertheless, due to time
constraints, we use a less conservative thinning threshold than the one
presented in \cite{Pierog:2015icrc}, which may be source of some uncertainties
in our results.

\begin{figure}[tbp]
\begin{minipage}[t]{0.49\textwidth} \postscript{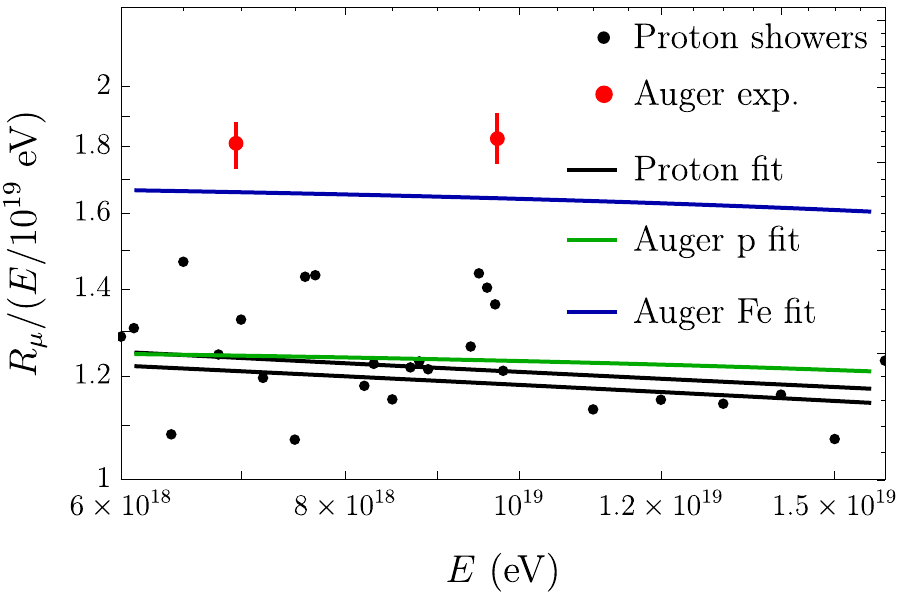}{0.99}
\end{minipage} \hfill \begin{minipage}[t]{0.49\textwidth}
\postscript{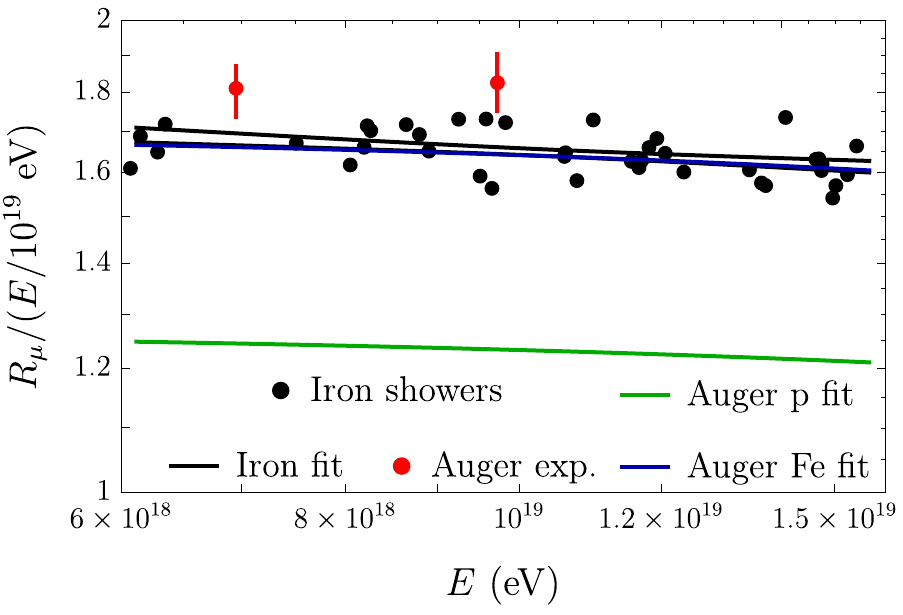}{0.99} \end{minipage}
\caption{Simulated showers (proton at left, iron at right) and
  $68\,\%$ C. L. fit bands, Auger fits to proton and iron simulations,
  and Auger measurements. All simulations using QGSJET II-04.}
\label{fig:RvsE}
\end{figure}

In order to measure the reliability of our simulations, we perform an analysis
of the muon content at ground level with the parameters suggested in
\cite{Aab:2014pza} for comparison. We concentrate on the energy interval
$10^{9.9}\lesssim E/\mathrm{GeV}\lesssim 10^{10.2}$, and simulate proton showers
at $67^\circ$ zenith angle. Following \cite{Aab:2014pza} a fit to a
parametrization
\begin{equation} \left<R_\mu\right>=a(E/{10^{19}\,\mathrm{eV}})^b \label{eq:Rfit}\end{equation} is assumed.
Here, $R_\mu=N_\mu/N_{\mu,19}$, where $N_\mu$ the number of muons at the Auger
observation level above $0.3\,\mathrm{GeV}$ for any energy, and $N_{\mu,19}$ the
same quantity evaluated at $10^{19}\,\mathrm{eV}$. Fitting (\ref{eq:Rfit}) to
our data we obtain $a=1.24\pm0.05$ and $b=0.90\pm0.02$, in good agreement with
the values presented in \cite{Aab:2014pza}. This agreement can be well
appreciated in Fig. \ref{fig:RvsE}.

The next step is to perform the same kind of analysis using primary
particles that may create fireballs in the first interaction.  We take
iron (A = 56) as reference. A new fit of (\ref{eq:Rfit}) to data
gives $a=1.65\pm0.01$ and $b=0.94\pm0.02$. The data and fit can be
also seen in Fig. \ref{fig:RvsE}.

A Heitler model predicts the total number of muons in a shower to
follow \begin{equation} N_{\mu,{\rm
      tot}}=A\left(\frac{E/A}{E_c}\right)^\alpha,\label{eq:heitler}\end{equation}
where $E$ and $A$ are the primary energy and mass number, respectively, $E_c$ the critical energy for
$\pi^\pm$ to decay into $\mu^\pm$. If we assume $R_\mu\propto N_{\mu,{\rm
    tot}}$, the exponents $\alpha$ and $b$ in (\ref{eq:heitler})
and (\ref{eq:Rfit}) must be the same. Since $b$ is almost the same for proton and
iron primaries, $\left<R_\mu\right>_{\rm Fe}/\left<R_\mu\right>_{\rm p}\sim
56^{1-b}\approx0.89\pm0.07$, where we have used the value of $b$ from
\cite{Aab:2014pza}. Using our fit parameters we obtain \begin{equation}
  \left<\frac{\left<R_\mu\right>_{\rm Fe}}{\left<R_\mu\right>_{\rm
      p}}\right>\approx1.13\pm0.05,\end{equation} where the \emph{external} average is
taken to be the considered energy interval. Due to the approximate nature of the
previous estimations, the numbers can be considered to be in agreement, and this
serves as another validation of the overall good behavior of our analysis with
the selected thinning.

\begin{figure}[tbp]
\begin{minipage}[t]{0.42\textwidth} \postscript{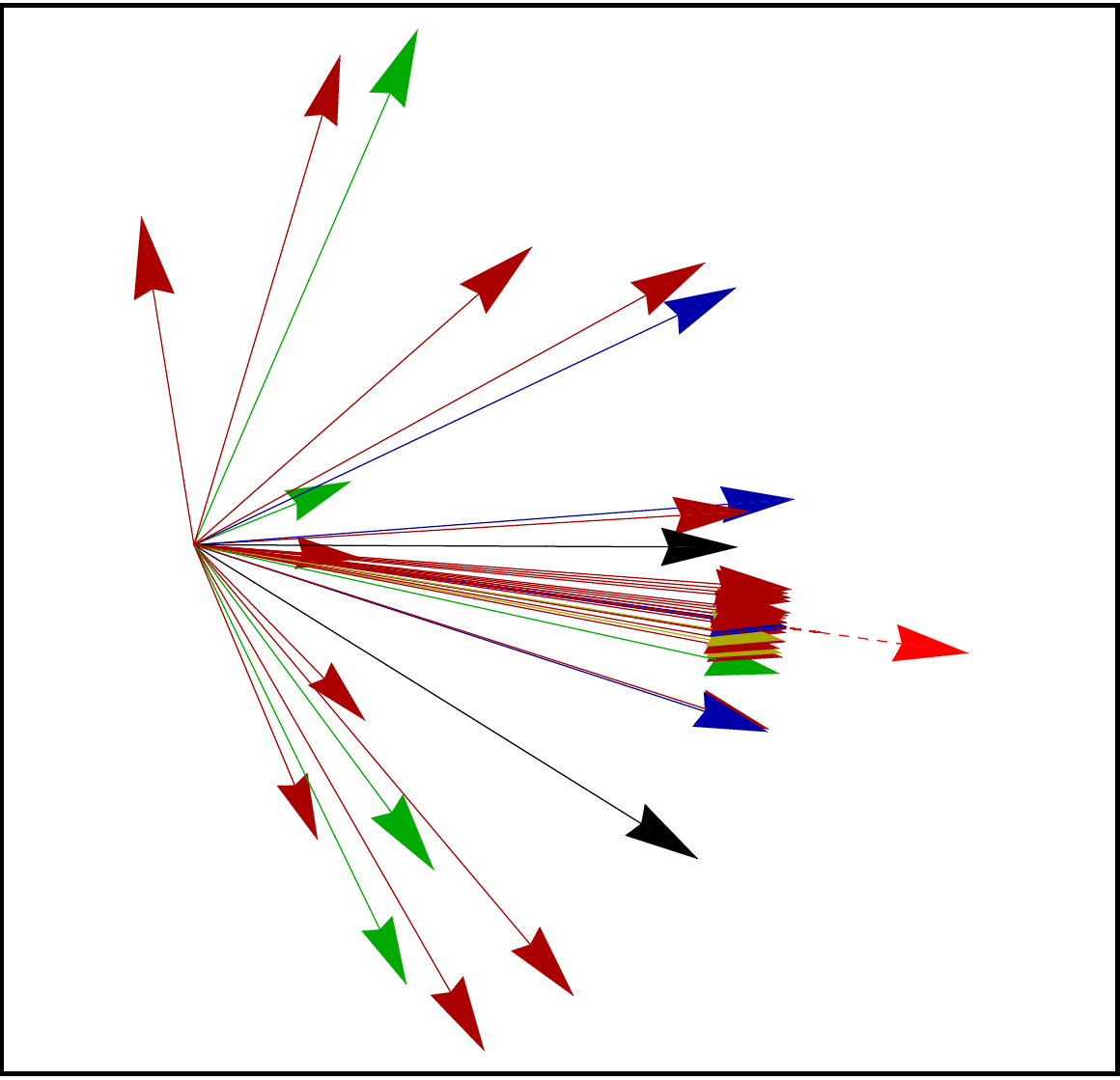}{0.99}
\end{minipage} \hfill \begin{minipage}[t]{0.58\textwidth}
\postscript{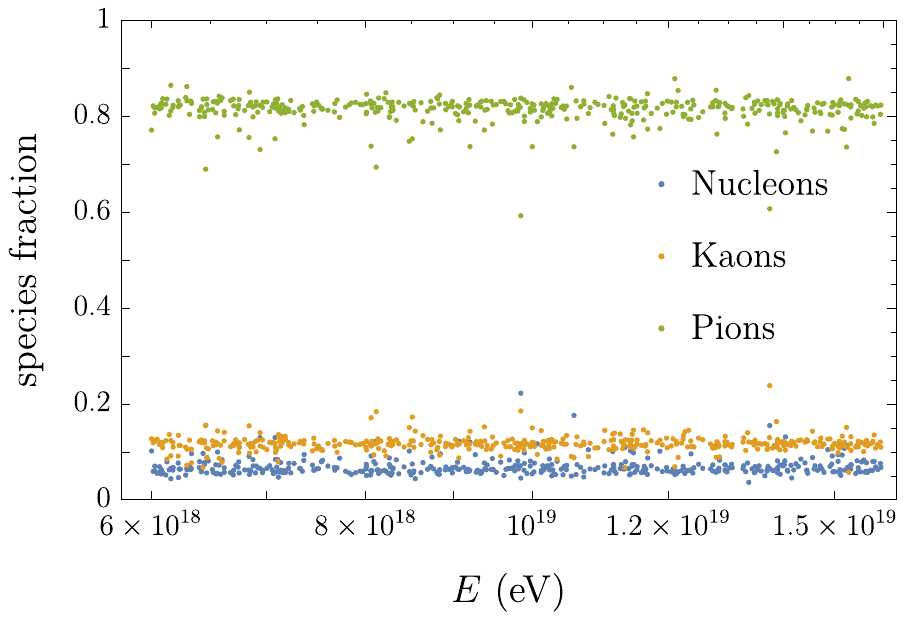}{0.99} \end{minipage}
\caption{{\em Left:} Secondaries normalized momenta in a N-N interaction in the target rest frame. Dashed red arrow points to shower axis. Colors represent particle species (green = nucleons, red = pions, blue = kaons, yellow=photons, black = others). The transverse  {\em Right:} fraction of kaons, pions and nucleons after the first interaction.}
\label{fig:first_int}
\end{figure}

Now, we proceed to the simulation and analysis of individual first
interactions. Using CRMC to interface with QGSJET II-04 we produce
iron-nitrogen collisions in the energy interval of our interest. We extract
the different particles created in the interaction, together with their
energies, momenta and child particles (in the case of the prompt decay of
neutral pions and kaons). Since we are precisely interested in those neutral
mesons, we revert those decays before performing any analysis. An example of the
secondary particles created in those interactions is shown in
Fig. \ref{fig:first_int}.

On the right panel of Fig. \ref{fig:first_int} we extract the information about
the fraction of particles of each kind, considering nucleons, pions and
kaons. As expected, the number of pions outnumbers that of kaons. As a first
approximation to see if the fireball model produces effects in the adequate
direction to explain current data, we invert kaon and pion
populations. According to Fig. \ref{fig:first_int}, the proportion of kaons will
be around $80\%$, much higher than the one of pions. This is, as stated previously, a first
approach to the effect that one would expect from a fireball model.

\begin{figure}[tbp]
\begin{minipage}[t]{0.49\textwidth} \postscript{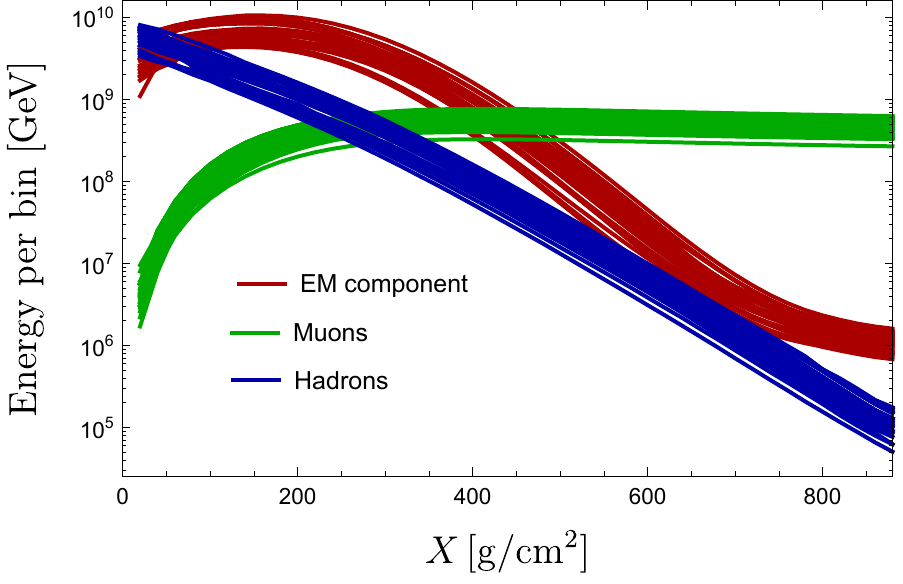}{0.99}
\end{minipage} \hfill \begin{minipage}[t]{0.49\textwidth}
\postscript{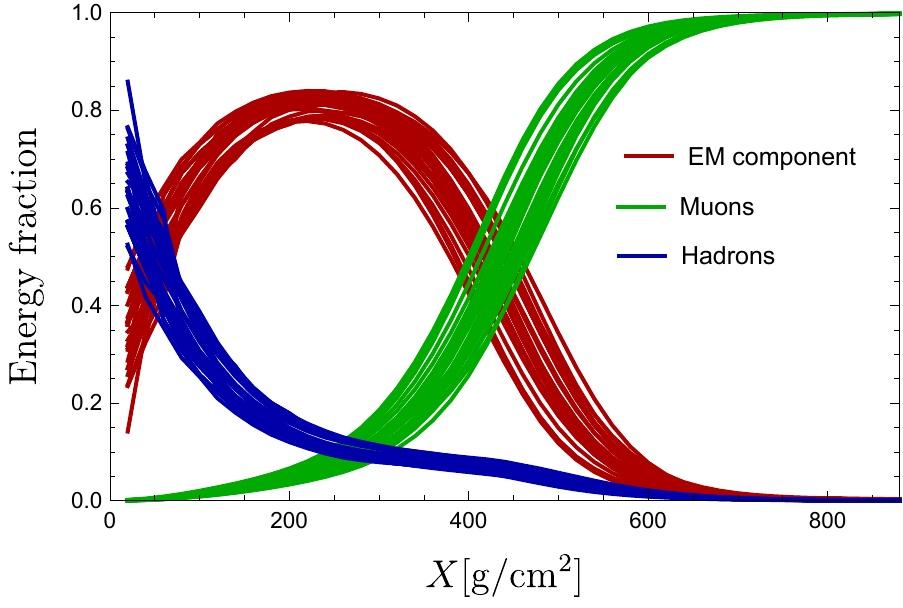}{0.99} \end{minipage}
\caption{Energy distribution and fraction of total energy for proton showers.}
\label{fig:edist_p}
\end{figure}

\begin{figure}[tbp]
\begin{minipage}[t]{0.49\textwidth} \postscript{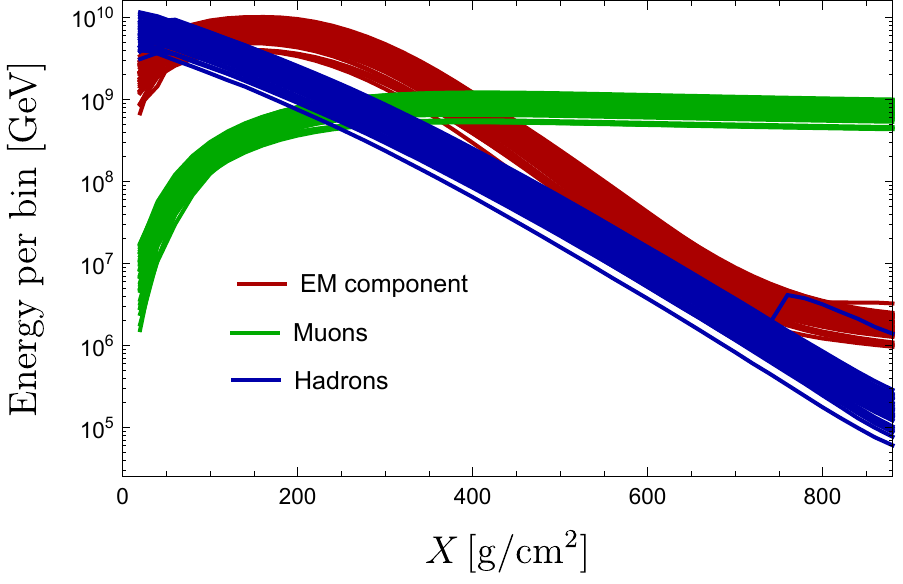}{0.99}
\end{minipage} \hfill \begin{minipage}[t]{0.49\textwidth}
\postscript{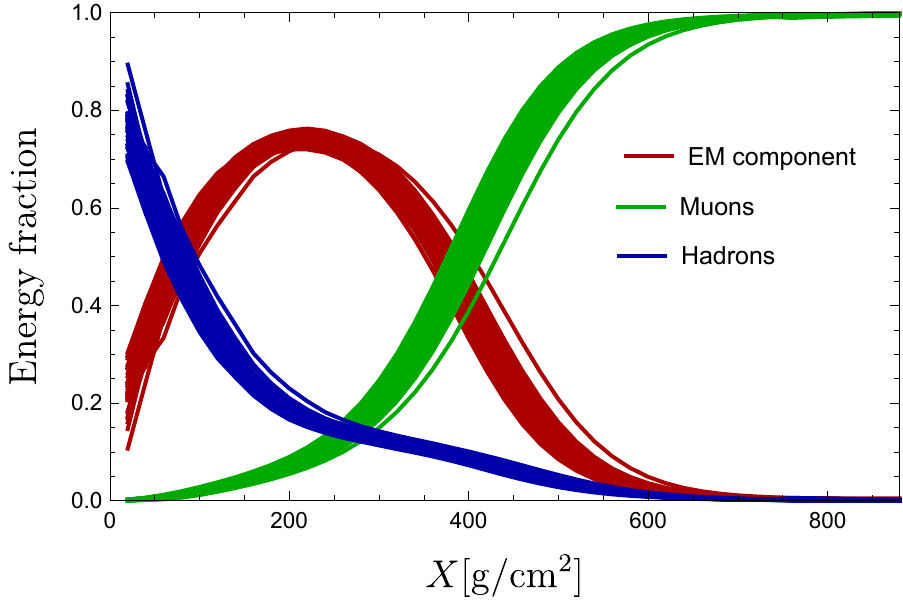}{0.99} \end{minipage}
\caption{Energy distribution and fraction of total energy for iron showers.}
\label{fig:edist_n}
\end{figure}

\begin{figure}[tbp]
\begin{minipage}[t]{0.49\textwidth} \postscript{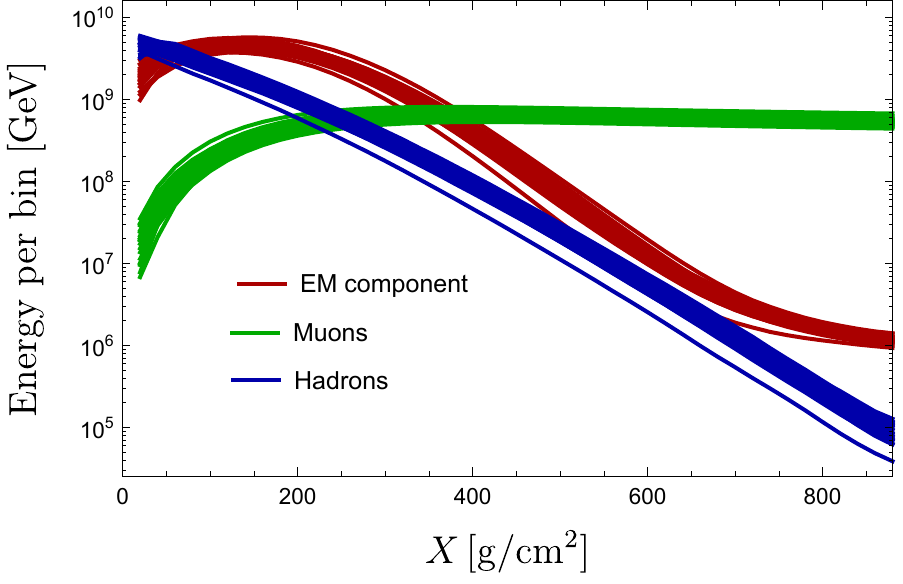}{0.99}
\end{minipage} \hfill \begin{minipage}[t]{0.49\textwidth}
\postscript{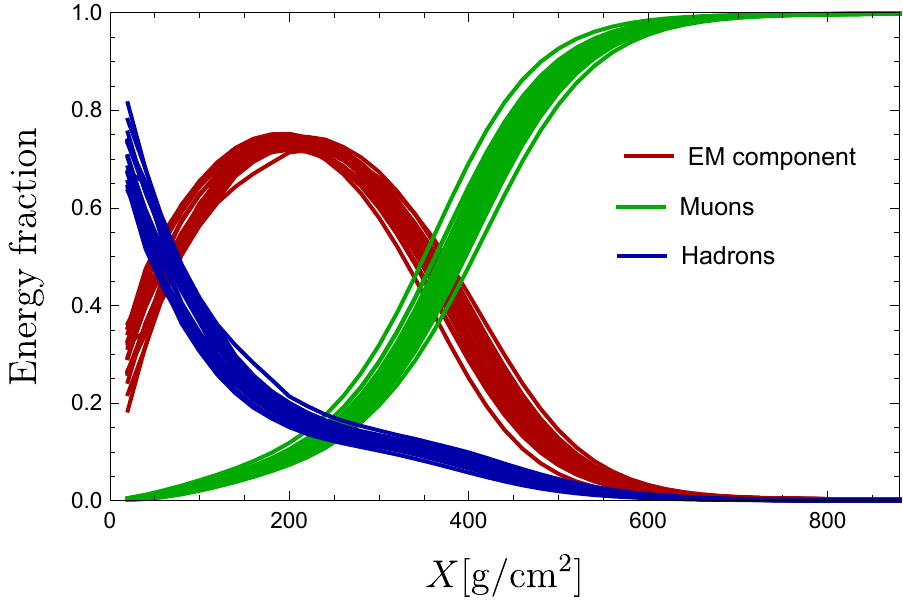}{0.99} \end{minipage}
\caption{Energy distribution and fraction of total energy for modified iron showers.}
\label{fig:edist_th}
\end{figure}

In order to compare the evolution of the shower in the different
situations, we show (Figs. \ref{fig:edist_p}, \ref{fig:edist_n} and
\ref{fig:edist_th}) the energy distribution and fraction of total
energy for the electromagnetic, muonic and hadronic components of the
different showers. One can see how the energy in the muonic component
increases for iron showers, as commented before. Note that the
range of energies for the showers in Fig. \ref{fig:edist_th} is much
smaller than the used in \ref{fig:edist_p} and
\ref{fig:edist_n}. 

\begin{figure}[tbp]
\postscript{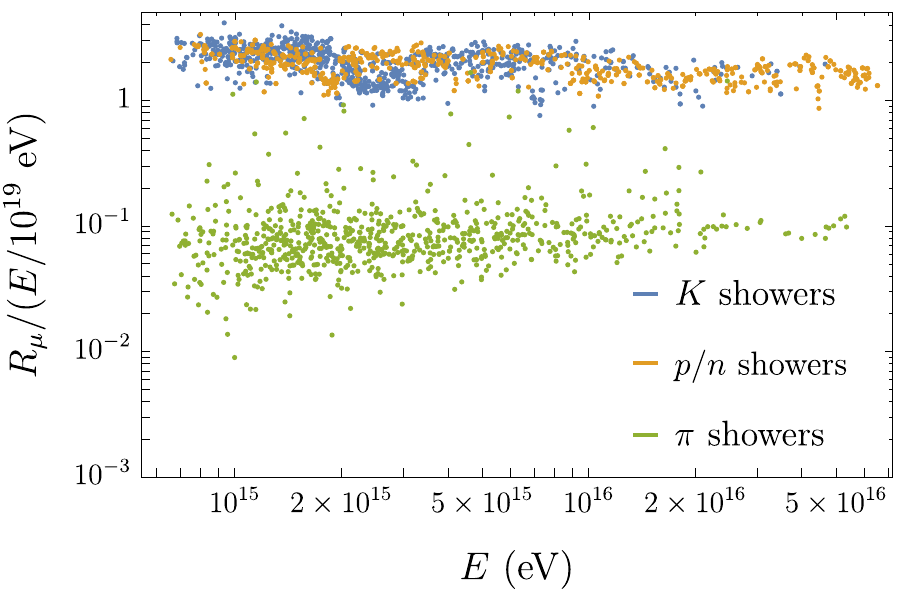}{0.69}
\caption{Energy distribution and fraction of total energy for modified iron showers.}
\label{fig:subshowers}
\end{figure}

In order to see more precisely how increasing the amount of energy in the kaon sector after the first interaction would increase the number of muons at ground, one can study kaon, pion and proton showers in the typical energy range of the products coming from primaries in the studied energy range. In Fig. \ref{fig:subshowers} we show the number of muons at ground for those kind of showers. While proton/neutron showers provide a similar amount of muons than kaon showers, pion showers remain around one order of magnitude lower. Then, transferring the energy contained in pions to kaons would produce the expected effect of increasing $R_\mu$.

Now, if the shower is initiated in a fireball explosion all the energy
of the incoming nucleus is redistributed among fundamental quarks and
gluons. This is a completely inelastic process, which differs from the
usual inelastic processes in that the fireball creates a higher
multiplicity of quarks and gluons and to a first approximation equally
partitions energy among the secondaries (thereby negating a large
leading particle effect). Thus, to simulate a complete fireball
explosion using inelastic collisions from LHC-tuned hadronic event
generators, we still have to account for the energy distribution
carried by the secondary nuclei produced in the primary collision. The
simplest and straightforward model of a fireball shower is then as
follows. Using CRMC to interface with QGSJET II-04 we produce
iron-nitrogen collisions in the energy interval of our interest. We
scan the secondaries searching for residual nuclei. If any nucleus is
found, we proceed to delete this nucleus from the list of secondaries
of the primary interaction. The hadronic collision package QGSJET
II-04 is call for separately, with the residual nuclear fragments as
projectiles. The secondaries produced by QGSJET II-04 in each of the
residual nucleus collisions are appended to the original list of
secondaries. We scan repeatedly the list of secondaries duplicating
the procedure detailed above until no more residual nuclei are
found. Then we search for pions and kaons in the final list of
secondaries. We change each pion whose kinetic energy is larger than
$m_K - m_\pi$ onto a kaon retaining charge, i.e. $\pi^+ \to K^+$,
$\pi^- \to K^-$, $\pi^0 \to K_l^0$ or $K_s^0$ (each case at random
with 50\% probability), and each kaon onto a pion also retaining
charge, i.e. $K^+ \to \pi^+$, $K^− \to \pi^-$, $K_l^0 \to \pi^0$,
$K_s^0 \to \pi^0$. We always preserve the total (rest + kinetic)
energy and particle direction of motion. Interestingly, UHECR showers
simulated following all of the previous considerations give an increment in $R_\mu $ by
${\cal O}(10\%)$. We conclude that to a first approximation fireball
induced showers would accommodate Auger observations shown in
Fig.~\ref{fig:RvsE}. A precise determination of the increase in
$R_\mu$ that would follow from a given modification of the particle
content after the first interaction is still ongoing and will be
presented elsewhere.

\section*{Acknowledgments}

We would like to acknowledge many useful discussions with Carlos
Garc\'{\i}a Canal, Sergio Sciutto, and our colleagues of the Pierre
Auger Collaboration.  This work has been partially supported by the
U.S. National Science Foundation (NSF) Grant No. PHY-1460394 (JFS and
LAA), by the National Aeronautics and Space Administration (NASA)
Grant No. NNX13AH52G (LAA and TCP), and by the U. S. Department of
Energy (DoE) Grant No. DESC-0011981 (TJW).

\end{document}